\documentclass[reprint,amsmath,amssymb,nofootinbib,aps,prc]{revtex4-1}
\usepackage{graphicx}
\usepackage{multirow}
\usepackage{relsize}
\newcommand{\snn} {\sqrt{s_{_{\rm NN}}}}
\newcommand{\lefttwo}{\!\!}
\newcommand{\leftfive}{\!\!\!\!\!\!}

\begin{document}

\title{Calculating the initial energy density in heavy ion collisions
  by including the finite nuclear thickness}
\author{Todd Mendenhall}
\email[]{mendenhallt16@students.ecu.edu}
\author{Zi-Wei Lin}
\email[]{linz@ecu.edu}
\address{Department of Physics, East Carolina University, Greenville, NC
27858, USA}
\date{\today}

\begin{abstract}

The initial energy density produced in heavy ion collisions can
be estimated with the Bjorken energy density formula after choosing a
proper formation time $\tau{_{\rm F}}$. However, the Bjorken formula breaks
down at low energies because it neglects the finite nuclear
thickness. Here we include both the finite time duration and finite
longitudinal extension of the initial energy production. 
When $\tau{_{\rm F}}$ is not too much smaller than the
crossing time of the two nuclei, our results are similar to those from
a previous study that only considers the finite
time duration. In particular, we find that at low energies the initial
energy density has a much lower maximum value but
evolves much longer than the Bjorken formula, while at large-enough
$\tau{_{\rm F}}$ and/or high-enough energies our result approaches the Bjorken
formula. We also find a qualitative difference in that our maximum
energy density $\epsilon^{\rm max}$ at $\tau{_{\rm F}}=0$ is finite, while the
Bjorken formula  diverges as  $1/\tau{_{\rm F}}$ and the previous result
diverges as $\ln (1/\tau{_{\rm F}})$ at low energies but as $1/\tau{_{\rm F}}$ at 
high energies.  Furthermore, our solution of the energy density
approximately satisfies a scaling relation. As a result, the
$\tau{_{\rm F}}$-dependence of $\epsilon^{\rm max}$ determines the
$A$-dependence, and the weaker $\tau{_{\rm F}}$-dependence of $\epsilon^{\rm 
  max}$ in our results at low energies means a slower increase of
$\epsilon^{\rm max}$ with $A$.
\end{abstract}

\maketitle

\section{Introduction}

The quark-gluon plasma (QGP) has been created in relativistic heavy ion
collisions 
\cite{Gyulassy:2004zy,Arsene:2004fa,Back:2004je,Adams:2005dq,Adcox:2004mh}. 
In the study of QGP properties, a key variable is the energy
density produced in such collisions. 
The maximum value and time evolution 
of the produced energy density affect the trajectory of an event 
on the temperature-baryon chemical potential plane. 
For lower collision energies such as those in the Beam Energy Scan
program at RHIC \cite{Mohanty:2011nm,Luo:2017faz,Adamczyk:2017iwn,
Keane:2017kdq}, 
the event trajectories relative to the location of the possible QCD
critical point \cite{Stephanov:2011pb,Bzdak:2019pkr} 
could significantly affect the experimental observables
and their sensitivities to the critical point
\cite{Stephanov:2011pb,Li:2018ygx}. 
For hydrodynamic models, the initial energy density 
including its spatial and temporal dependences 
\cite{Okai:2017ofp,Shen:2017ruz,Du:2018mpf} is an essential input for
the subsequent hydrodynamical evolution of the dense matter.

The Bjorken energy density formula \cite{Bjorken:1982qr} 
is a convenient way to estimate the initial energy density averaged
over the transverse area of a relativistic heavy ion collision:
\begin{equation}
\epsilon_{Bj}(t)=\frac{1}{t \, A_{\rm T}}\frac{dE_{\rm T}}{dy}.
\label{bjorken}
\end{equation}
In the above, $A_{\rm T}$ is the transverse overlap area of the two nuclei,
and $dE_{\rm T}/dy$ is the transverse energy rapidity density at
mid-rapidity (for estimating the initial energy density in the central
region), which is often taken as the experimental $dE_{\rm T}/dy$
value in the final state. 
Because this formula diverges as $t \to 0$, one must choose
a non-zero initial time, usually by assuming a finite formation time
$\tau{_{\rm F}}$ for the produced particles. 
Note that the Bjorken energy density formula
assumes that all initial particles are produced at $t=0$ and $z=0$
before they start to propagate and later become on-shell. Therefore it
is valid at high energies where the Lorentz-contracted nuclear
thickness is negligible compared to the formation time, while it is
expected to break down at low energies when the finite nuclear
thickness becomes comparable to or larger than the formation time
\cite{Adcox:2004mh}. 
For central nucleus-nucleus collisions, it takes the following finite
time in the hard sphere model of the nucleus for two 
identical nuclei of mass number $A$ to completely cross each other in
the center-of-mass frame:
\begin{equation}
d_t=\frac{2R_A}{{\rm \sinh}\, y_{cm}},
\label{crossing time}
\end{equation}
where $y_{cm}$ is the rapidity of the projectile nucleus. 
For central Au+Au collisions at $\snn=50$ GeV, for example, $d_t \approx
0.5$ fm/$c$ is comparable to the usual value of the parton
formation time when we take $R_A=1.12 A^{1/3}$ fm as the nuclear
radius. Therefore we may expect the Bjorken formula to break down for
central Au+Au collisions at $\snn \lesssim 50$ GeV \cite{Lin:2017lcj}.

A previous study by one of us \cite{Lin:2017lcj} 
extended the Bjorken energy density formula by considering that the
initial energy is produced over a finite duration time $[0,d_t]$. 
Its analytical result approaches the Bjorken formula 
at high energies. At low energies, however, it finds that the
maximum energy density $\epsilon^{\rm max}$ reached is much lower but 
the time evolution of the energy density 
(e.g., as measured by the time duration when the energy density stays
above $\epsilon^{\rm max}$/2) is much longer  
in comparison with the Bjorken formula. 
In addition, the maximum energy density in the low-energy limit
depends on $\ln (1/\tau{_{\rm F}})$, therefore at low energies it is much less
sensitive to the uncertainty of the formation time than the Bjorken
formula, which energy density depends on $1/\tau{_{\rm F}}$. 

However, the analytical method of the previous study
\cite{Lin:2017lcj} did not take into account the finite longitudinal
width (in $z$) of the initial energy production. In this work we 
include both the finite duration time and the finite
$z$-width of the initial energy production. 
We then study the time evolution of the produced initial energy density in
the central spacetime-rapidity region (i.e., $\eta_s \approx 0$) in the
center-of-mass frame of central collisions of two identical nuclei. 
Note that as in the previous study we neglect subsequent interactions
among the produced particles, which can be modeled by transport models 
\cite{Xu:2004mz,Lin:2014tya} or hydrodynamic models 
\cite{Okai:2017ofp,Shen:2017ruz}; we only study the energy produced
from primary collisions between nucleons from the projectile and
target nuclei.

\section{Method}

\begin{figure}
\centering
\includegraphics[width=\linewidth]{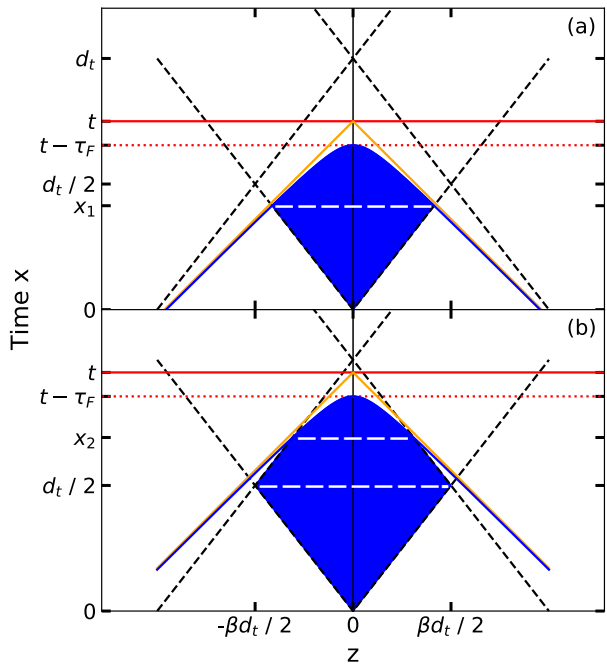}
\caption{Schematic diagram for the crossing of two identical
  nuclei, where partons can be produced anywhere inside the
  rhombus, for (a) the first and (b) the second piecewise solution in
  Table I. The solid diagonal lines
  represent the light cone boundaries for partons that can reach $z
  \approx 0$ at time $t$, while the hyperbola represents the 
  boundary of these partons after considering the  
  formation time $t_F=\tau{_{\rm F}}\,{\rm \cosh}\,y$.}
\label{fig:crossing-diagram}
\end{figure}

We begin by examining in Fig.~\ref{fig:crossing-diagram} 
the crossing of two identical relativistic nuclei traveling  along
the $\pm z$ directions with speed $\beta={\rm \tanh}\,y_{cm}$. 
As the nuclei cross each other, the full $z$-width of the overlap
region first increases from $0$ at $t=0$ to 
$\beta d_t$ at $t=d_t/2$ and then decreases back to $0$ at $t=d_t$. 
We refer to this rhombus (the area surrounded by the four dashed
lines) as the production area
because it covers the area of primary collisions in the $z-t$
plane \cite{Kajantie:1983ia,Spieles:1999kp}. For simplicity, in this
study we neglect the transverse expansion of the overlap volume as 
well as the slowing down of participant nucleons during the
primary collisions, as done in the Bjorken energy density formula
\cite{Bjorken:1982qr} and the previous extension study
\cite{Lin:2017lcj}.

We are interested in the initial energy in the narrow region $z \in
[-d,d]$ within the transverse overlap area $A_{\rm T}$ at time $t$. 
An initial parton can be produced at $z$-coordinate $z_0$ and time
$x$, i.e., at point $(z_0,x)$, within the production area, and it is
then assumed to propagate with its velocity until it is formed 
after a formation time $t_F$. 
For a parton produced at time $x$ to be within the narrow range
$z \in [-d,d]$ at time $t$, its rapidity $y$ must satisfy the following
condition:
\begin{equation}
\frac{-d-z_0}{t-x} \leq {\rm \tanh}\,y \leq \frac{d-z_0}{t-x}.
\label{rapidity range}
\end{equation}
Therefore, in the limit $d \to 0$ the rapidity and its allowed range become
\begin{equation}
y \to y_0=  {\rm tanh}^{-1} \left (\frac{-z_0}{t-x} \right ),
~\Delta y = \frac{2d~{\rm \cosh}^2 y}{t-x}.
\label{y0}
\end{equation}
So the initial energy density averaged over the transverse area at
time $t$ is
\begin{equation}
\begin{split}
\epsilon (t)&=\frac{1}{2dA_{\rm T}}\iint_S dx dz_0 
{\frac{d^3m_{\rm T}}{dx dz_0 dy}}  \Delta y~{\rm \cosh}\,y\\
&=\frac{1}{A_{\rm T}}\iint_S \frac{dx dz_0}{t-x}
\frac{d^3m_{\rm T}}{dx dz_0 dy_0} {\rm \cosh}^3 y_0.
\end{split}
\label{epsilon}
\end{equation}
In the above, $m_{\rm T}$ is the transverse mass that is the same as
the transverse energy $E_{\rm T}$ at $y=0$. We use the notation
$m_{\rm T}$ in this study to differentiate our $dm_{\rm T}/dy$ from
the experimentally measured $dE_{\rm T}/dy$ from
the PHENIX Collaboration \cite{Adler:2004zn}.

The limits of integration in Eq.\eqref{epsilon} that 
determine the integration area $S$ depend on time $t$. 
First, any $(z_0,x)$ point needs to be within
the production area, shown in Fig.~\ref{fig:crossing-diagram} as the
diamond-shaped area formed by the four dashed lines in each panel. 
Secondly, the light cone limits the production points of allowed
partons  below the two diagonal solid lines in each panel of 
Fig.~\ref{fig:crossing-diagram}. Finally, a parton needs to be formed
by time $t$ due to its finite formation time.
Now we take the formation time of a parton in
the center-of-mass frame as 
\begin{equation}
t_F=\tau{_{\rm F}}~{\rm \cosh}\,y,
\end{equation}
i.e., a proper formation time $\tau{_{\rm F}}$ multiplied by a
time-dilation factor. For a parton produced at point $(z_0,x)$ that
would reach point $(\approx\!\!0,t)$ and contribute to Eq.\eqref{epsilon},
its formation time is $\tau{_{\rm F}} \, {\rm \cosh}\,y_0$.
Therefore any allowed production point needs to be below a 
formation time hyperbola, which is given by 
\begin{equation}
x=t-\sqrt {z^2+\tau{_{\rm F}}^2}.
\label{tauFcurve}
\end{equation}
Note that for finite $\tau{_{\rm F}}$ this formation time requirement is
always stricter than the light cone requirement, while for
$\tau{_{\rm F}}=0$ the hyperbola reduces to the light cone boundaries.

Since the integration limits of $(z_0,x)$ in Eq.\eqref{epsilon} 
depend on time, our solution of the energy density $\epsilon(t)$ 
is a piecewise function in time. 
We now consider a more general case than
Fig.~\ref{fig:crossing-diagram} in that the crossing of two nuclei
starts at time $t_1$ and ends at time $t_2$ and thus the rhombus
production area is bound by the $z=\pm \beta (x-t_1)$ and $z=\pm \beta 
(x-t_2)$ lines. Then we summarize the integration limits in
Table~\ref{general limits}, noting that $\epsilon(t)=0$ for $t \in
[0, t_1+\tau{_{\rm F}})$. In the table, $t_a$ is the observation time when the
formation time hyperbola intersects the two middle vertices of the
production area at $(z_0,x)=(\pm \beta t_{21}/2, t_{\rm mid})$:
\begin{equation}
t_a=t_{\rm mid}+\sqrt{\tau{_{\rm F}}^2+\left(\frac{\beta t_{21}}{2}\right)^2},
\label{ta}
\end{equation}
where we define
\begin{equation}
t_{21}=t_2-t_1, ~ t_{\rm mid}=(t_1+t_2)/2. 
\end{equation}
The first piecewise solution is for time $t \in [t_1+\tau{_{\rm F}},t_a)$, 
where the formation time hyperbola intersects the lower boundaries of
the production area, i.e., the $z=\pm \beta (x-t_1)$ lines, at time
$x_1$ that is given by 
\begin{equation}
x_i\!=\!\frac{t\!-\!\beta^2t_i\!-\!\sqrt{\beta^2\left [(t-t_i)^2\!-\!\tau{_{\rm F}}^2\right
    ]\!+\!\tau{_{\rm F}}^2}}{1-\beta^2}, {\rm with}~i\!=\!1,2. 
\label{x12}
\end{equation}
As shown in Fig.~\ref{fig:crossing-diagram}(a) and Table~\ref{general
  limits}, the first piece $\epsilon_1(t)$ has two integration areas:
a triangular area below time $x_1$ and another area under the hyperbola. 
For the latter area, the $z_0$-range is $[-z_F(x),z_F(x)]$, where 
$\pm z_F(x)$ are the $z$-coordinates of the formation time hyperbola 
at a given time $x$:
\begin{equation}
z_F(x)=\sqrt{\left(t-x\right)^2-\tau{_{\rm F}}^2}.
\label{proper time boundary}
\end{equation}
The second piecewise solution is for time $t \in [t_a,t_2+\tau{_{\rm F}})$, 
where the formation time hyperbola intersects the upper boundaries of
the production area, i.e., the $z=\pm\beta(x-t_2)$ lines, at time
$x_2$ as given by Eq.\eqref{x12}. 
As shown in Fig.~\ref{fig:crossing-diagram}(b), the second piece
$\epsilon_2(t)$ has three integration areas: the lower half of the
rhombus (a triangle), the upper half of the rhombus below time $x_2$
(a trapezoid), and the rhombus above time $x_2$ but under the
hyperbola curve. 
Note that in each panel of Fig.~\ref{fig:crossing-diagram}
the different integration areas are separated by the dashed 
line(s) inside the shaded full integration area.
Finally, the third piece $\epsilon_3(t)$ gives the 
solution for time $t \in [t_2+\tau{_{\rm F}}, \infty)$, where the
integration is over $(z_0,x)$ in the full rhombus.

If we neglect the finite time duration and longitudinal
width of the initial energy production and thus 
make the replacement $d^3m_{\rm T}/(dxdz_0dy) \to
\delta(z_0)\delta(x)dm_{\rm T}/dy$, we recover the Bjorken energy
density formula of Eq.\eqref{bjorken}. 
On the other hand, if we consider the finite time duration but neglect
the finite longitudinal width and thus make the replacement
 $d^3m_{\rm T}/(dxdz_0dy) \to \delta (z_0) d^2m_{\rm T}/(dxdy)$,
Eq.\eqref{epsilon} then reduces to the previously known solution:
Eq.(5) of Ref.\cite{Lin:2017lcj}. 
Note that $\epsilon(t)$ is higher for a smaller $\tau{_{\rm F}}$
(at given $\snn$, $A$ and $t$) because the integration area gets
bigger, except that the late-time $\epsilon(t)$ at $t > t_2+\tau{_{\rm F}}$
does not depend on $\tau{_{\rm F}}$.

\begin{table}
\begin{tabular}{|c|c|c|c|}
\hline
Piece & $t$ range & $x$ range & $z_0$ range\\
\hline
\multirow{2}{*}{$\epsilon_1(t)$} & \multirow{2}{*}{$[t_1+\tau{_{\rm F}},t_a)$}
& $[t_1,x_1)$ & $[-\beta(x-t_1),\beta(x-t_1)]$ \\ 
\cline{3-4}
& & $[x_1,t-\tau{_{\rm F}})$ & $[-z_F(x),z_F(x)]$ \\
\hline
\multirow{3}{*}{$\epsilon_2(t)$} & \multirow{3}{*}{$[t_a,t_2+\tau{_{\rm F}})$}
  & $[t_1,t_{\rm mid})$ &  $[-\beta(x-t_1),\beta(x-t_1)]$ \\ 
\cline{3-4}
& & $[t_{\rm mid},x_2)$ & $[-\beta(t_2-x),\beta(t_2-x)]$ \\
\cline{3-4}
& & $[x_2,t-\tau{_{\rm F}})$ & $[-z_F(x),z_F(x)]$ \\
\hline
\multirow{2}{*}{$\epsilon_3(t)$} & \multirow{2}{*}{$[t_2+\tau{_{\rm F}},\infty)$}
& $[t_1,t_{\rm mid})$ & $[-\beta(x-t_1),\beta(x-t_1)]$ \\   
\cline{3-4}
& & $[t_{\rm mid},t_2]$ & $[-\beta(t_2-x),\beta(t_2-x)]$ \\
\hline
\end{tabular}
\caption{Piecewise solution of $\epsilon(t)$ for different ranges
  of time $t$ together with the corresponding integration limits for
  the production time $x$ and production $z$-coordinate $z_0$.}
\label{general limits}
\end{table}

To proceed further, we now consider central Au+Au collisions
and specify the function $d^3m_{\rm T}/(dxdz_0dy)$ in Eq.\eqref{epsilon}.
We first assume that the initial transverse mass rapidity density of
produced partons per production area can be written in a factorized form: 
\begin{equation}
\frac{d^3m_{\rm T}}{dx dz_0~dy}=g(z_0,x)~\frac{dm_{\rm T}}{dy}.
\label{production rate}
\end{equation}
The area density function $g(z_0,x)$ is normalized as
\begin{equation}
\iint_{S_0} dx dz_0~g(z_0,x)=1
\end{equation}
so that $dm_{\rm T}/dy$ represents the initial rapidity density of
the transverse mass of all produced partons. 
We further make the simplest assumption that partons are produced
uniformly over the full production area $S_0$, i.e.,
\begin{equation}
g(z_0,x)=\frac{2}{\beta t_{21}^2}.
\label{g function}
\end{equation}

We parametrize the initial $dm_{\rm T}/dy$ of produced partons as
a Gaussian function in rapidity:
\begin{equation}
\frac{dm_{\rm T}}{dy}=
\frac{dm_{\rm T}}{dy}(0)~{\mathlarger e}^{^{-\frac{y^2}{2\sigma^2}}}, 
\label{dmtdyg}
\end{equation}
where we use the notation $F(0)$ to represent the value of $F(y)$ at
$y=0$. 
We then take the peak value of $dm_{\rm T}/dy$ at different collision
energies from a parametrization of the results from the string melting
version of the AMPT model \cite{Lin:2017lcj}: 
\begin{equation}
\frac{dm_{\rm T}}{dy}(0)=168 \left (\frac{\snn}{\rm GeV}-0.930
\right )^{0.348} {\rm~GeV}.
\label{dmtdy0}
\end{equation}
To determine the Gaussian width $\sigma$, we take advantage of the
conservation of energy by assuming that for central collisions all
incoming nucleons are participant nucleons:
\begin{equation}
\int\frac{dm_{\rm T}}{dy}~{\rm \cosh}\,y~dy=A\snn.
\label{energy-conservation}
\end{equation}
We then obtain
\begin{equation}
\sigma = \sqrt {W_0(r^2)}, {\rm ~with~} 
r=\frac{A\snn}{\sqrt{2\pi}{\frac{dm_{\rm T}}{dy}}(0)},
\label{sigma}
\end{equation}
where $W_0(x)$ is the $k=0$ branch of the Lambert $W$ function (or the
omega function) $W_k(x)$.
Finally, we can write the initial energy density 
averaged over the transverse area as 
\begin{equation}
\epsilon(t)=\frac{2}{A_{\rm T}\beta t_{21}^2} \frac{dm_{\rm T}}{dy}(0)
\iint_S \frac{dx dz_0}{t-x} 
{\mathlarger e}^{^{\!-\frac{y_0^2}{2\sigma^2}}} {\rm \cosh}^3 y_0.
\label{epsilon2}
\end{equation}

\begin{figure}
\centering
\includegraphics[width=\linewidth]{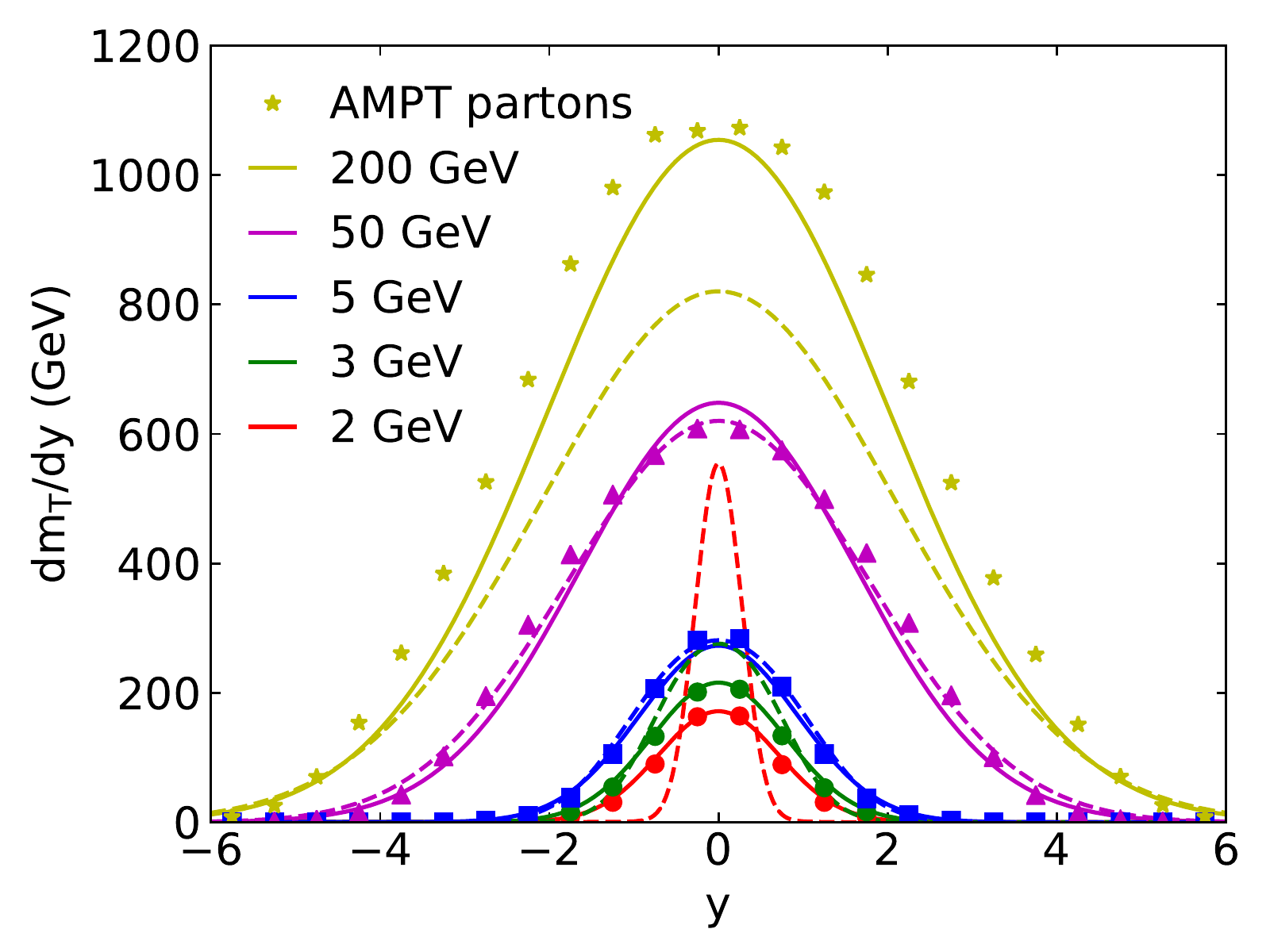}
\caption{Parametrized initial rapidity density of transverse mass of
  produced partons (solid curves) for central Au+Au collisions at
  $\snn$ = 2, 3, 5, 50, and 200 GeV. Symbols represent the results of
  initial partons from the AMPT model, while dashed curves represent
  the parametrized hadron $dm_{\rm T}/dy$ at these energies.}
\label{fig:dmtdy}
\end{figure}

Figure~\ref{fig:dmtdy} shows the $dm_{\rm T}/dy$ of produced partons 
as given by Eq.\eqref{dmtdyg} in central Au+Au collisions at several
energies (solid curves), where we see a monotonous increase of the
peak value and the Gaussian width with the collision energy. 
Symbols represent the results of initially produced partons from the string
melting version of the AMPT model \cite{Lin:2017lcj}, which show the
same qualitative features. Note that in more realistic calculations
such as those from the HIJING model
\cite{Wang:1991hta,Gyulassy:1994ew} or the AMPT model 
\cite{Lin:2004en} a small fraction of the incoming nucleons are
spectators in central collisions.

\section{Results for central Au+Au collisions}

Our results for $\epsilon(t)$ depend on choosing specific values 
for the time parameters $\tau{_{\rm F}}$, $t_1$, and $t_2$. 
As in the previous study \cite{Lin:2017lcj}, we take 
\begin{equation}
t_1=0.2\,d_t, ~t_2=0.8\,d_t,
\end{equation}
instead of the naive choice of 
$t_1=0$ and $t_2=d_t$; this is understandable because a boosted
nucleus has the shape of an ellipsoid instead of a uniform disk. 
These particular values are chosen \cite{Lin:2017lcj} so that the
width of the production time distribution is similar to the results
from the string melting version of the AMPT model.

\begin{figure}
\centering
\includegraphics[width=\linewidth]{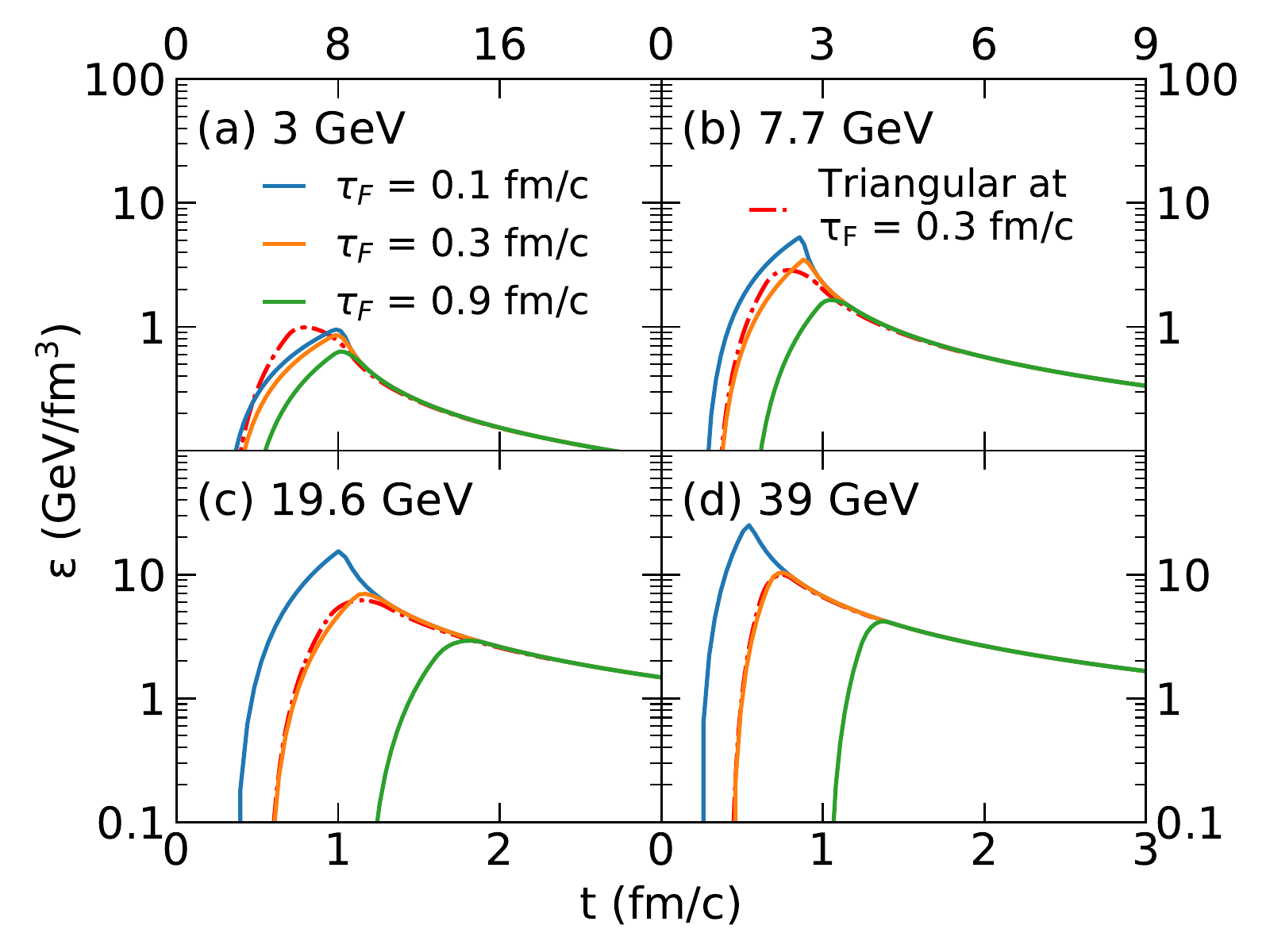}
\caption{Energy density of produced partons in central Au+Au
  collisions at $\snn$ = 3, 7.7, 19.6, and 39 GeV for $\tau{_{\rm F}}=0.1, 0.3$
  and $0.9$ fm/$c$; the triangular solution for $\tau{_{\rm F}}=0.3$ fm/$c$
  is also shown for comparison.}
\label{fig:epsilon-vs-t}
\end{figure}

Figure~\ref{fig:epsilon-vs-t} shows our results 
of the initial energy density versus time for central Au+Au
collisions  at $\snn$ = 3, 7.7, 19.6, and 39 GeV in four panels  
for several different $\tau{_{\rm F}}$ values. 
We see that the energy density first increases smoothly with time 
and that the late-time decrease is essentially the same for different
$\tau{_{\rm F}}$ values. 
In addition, the peak energy density increases with the decrease of
$\tau{_{\rm F}}$, but the relative increase is smaller at lower energies. 
These features are the same as those from the previous study that only
includes the finite time duration \cite{Lin:2017lcj}. 
Also, our results for $\tau{_{\rm F}}=0.3$ fm/$c$ are quite
close to those from the previous triangular time profile that took the
same $t_1$ and $t_2$ values  \cite{Lin:2017lcj}. 
This may be expected because the assumption in Eq.\eqref{g function}
of a uniform distribution in $(z_0,x)$ leads to a triangular time
profile in $x$ (after integrating over $z_0$). Note that the
triangular solution is also a piecewise solution 
\cite{Lin:2017lcj}: 
\begin{eqnarray}
\epsilon_{\rm tri} (t)&& =
\frac{4}{A_{\rm T} t_{21}^2} \frac{dm_{\rm T}}{dy}(0) \lefttwo
\left [\! -t \!+\! t_1 \!+\! \tau{_{\rm F}} \!+\! (t-t_1) \ln \! \left (\!
\frac{t-t_1}{\tau{_{\rm F}}} \! \right ) \right ] \!, \nonumber \\ 
&&{~\rm for~} t \in [t_1+\tau{_{\rm F}},t_{\rm mid}+\tau{_{\rm F}}); \nonumber \\ 
&&\leftfive=\frac{4}{A_{\rm T} t_{21}^2} \frac{dm_{\rm T}}{dy}(0) \lefttwo
\left [ t-t_2-\tau{_{\rm F}} +(t-t_1) \ln \! \left ( \!
    \frac{t-t_1}{t-t_{\rm mid}} \! \right ) \right . \nonumber \\ 
&& \left . + (t_2-t) \ln \! \left ( \! \frac{t-t_{\rm mid}}{\tau{_{\rm F}}}
    \! \right ) \right ] \!, \lefttwo {~\rm for~} t \in [t_{\rm mid} \!+\!
\tau{_{\rm F}},t_2 \!+\! \tau{_{\rm F}}); \nonumber \\ 
&&\leftfive=\frac{4}{A_{\rm T} t_{21}^2} \frac{dm_{\rm T}}{dy}(0) \lefttwo
\left [ (t-t_1) \ln \! \left ( \! \frac{t-t_1}{t-t_{\rm mid}} \!
   \right )  \right . \nonumber \\  
&& \left . +(t_2-t) \ln \! \left ( \! \frac{t-t_{\rm mid}}{t-t_2} \!
   \right ) \right ] \!, {~\rm for~} t \in [t_2+\tau{_{\rm F}},\infty).
\label{ettri}
\end{eqnarray}
Note that $dm_{\rm T}/dy(0)$ appears in the above solution because
only partons at $y \approx 0$ can enter the central spacetime-rapidity 
region of $\eta_s \approx 0$ when the finite $z$-width of the initial
energy production is neglected.

\begin{figure}
\begin{minipage}[h]{0.97\linewidth}
\includegraphics[width=1.\textwidth]{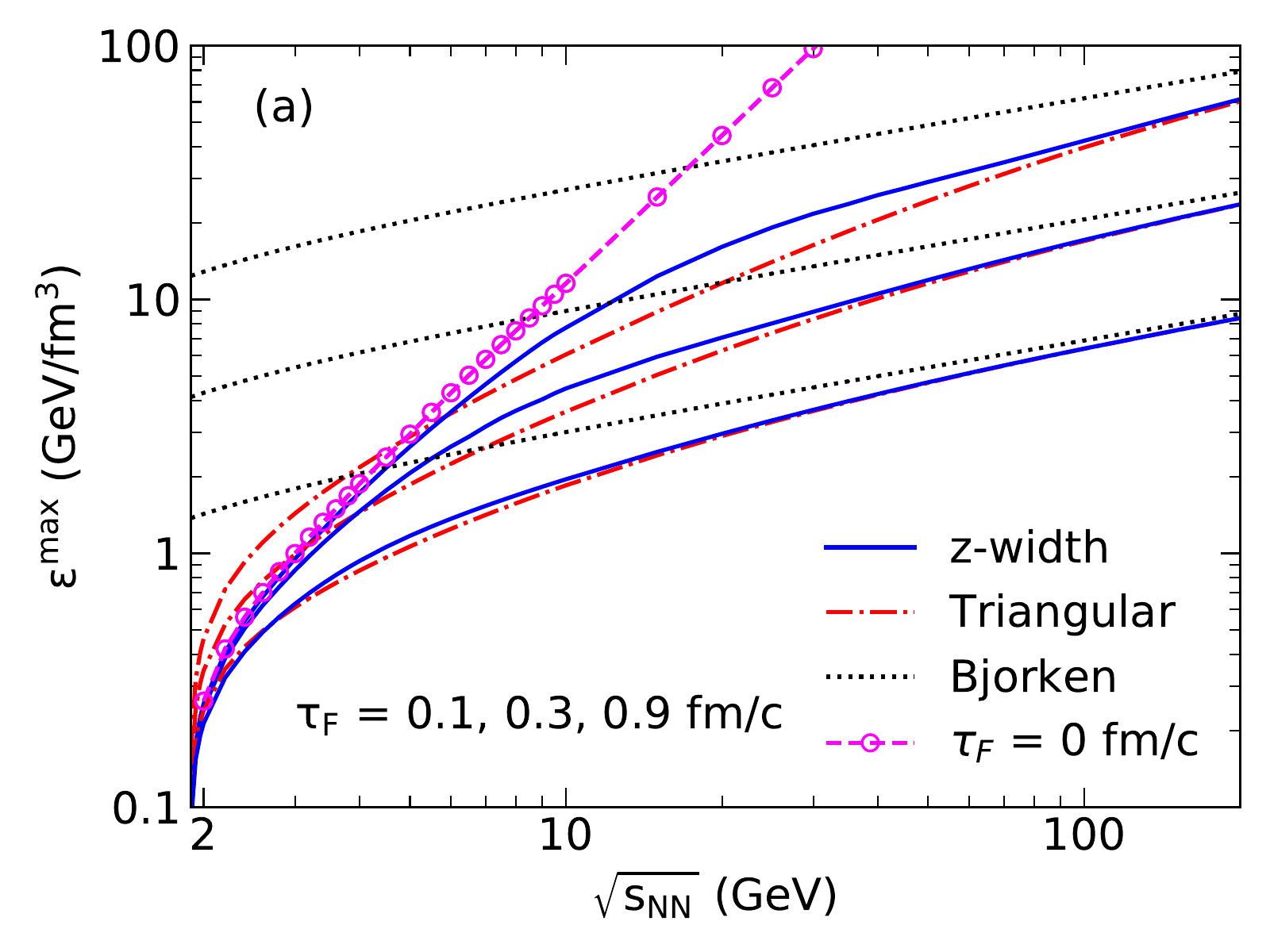}
\end{minipage}
\hfill
\begin{minipage}[h]{\linewidth}
\includegraphics[width=1.\textwidth]{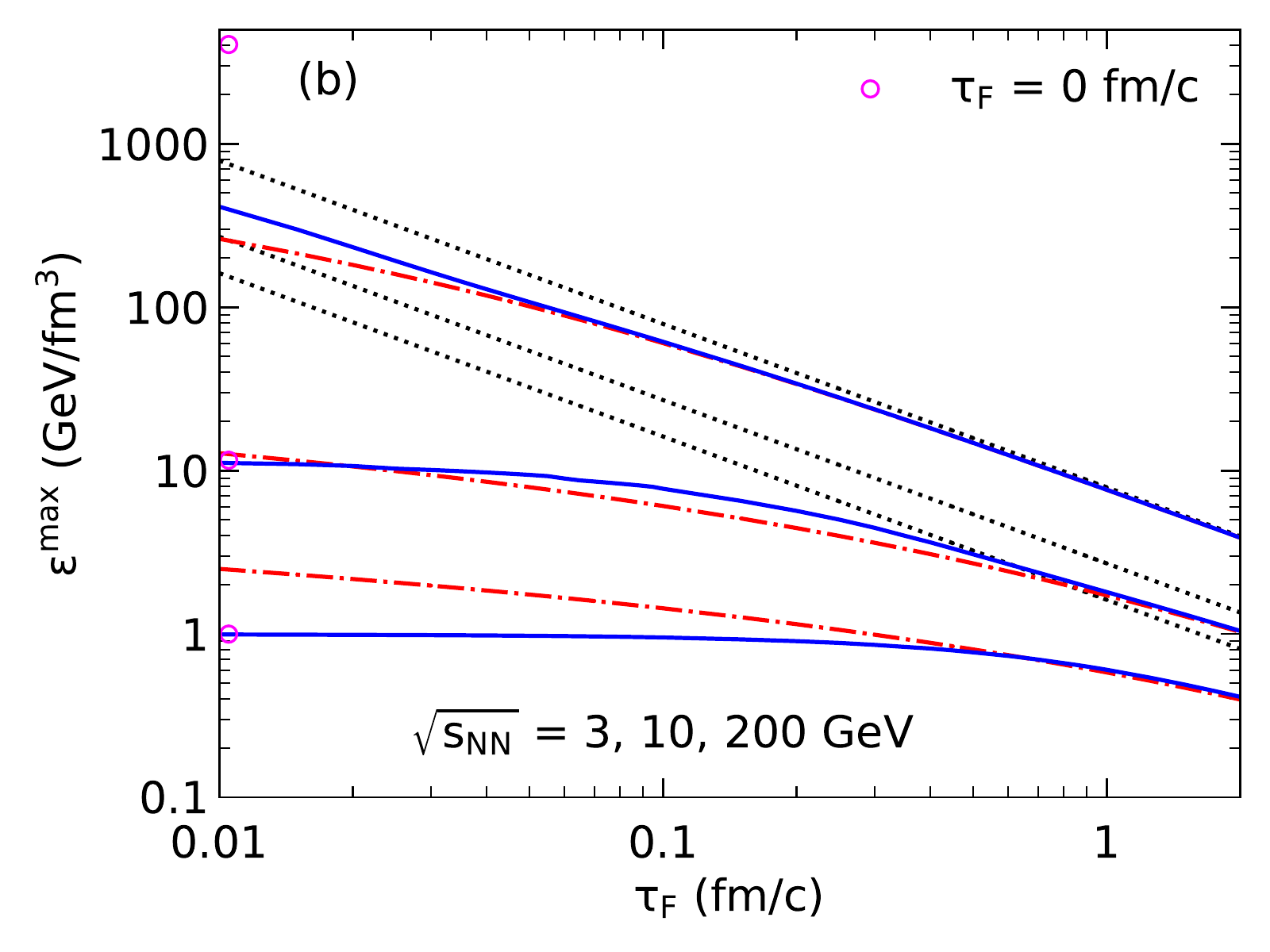}
\end{minipage}
\caption{Maximum energy density for central Au+Au collisions 
(a) as a function of collision energy at $\tau{_{\rm F}}=$ 0.1, 0.3 and 0.9
fm/$c$ and (b) as a function of proper formation time at $\snn=$ 3,
10, 200 GeV from our method, the triangular time profile, and the
Bjorken formula. Circles represent results for $\tau{_{\rm F}}=0$.} 
\label{fig:emax-snn-tauf}
\end{figure}

From each $\epsilon(t)$ curve we extract the maximum energy
density $\epsilon^{\rm max}$, whose values are shown 
in Fig.~\ref{fig:emax-snn-tauf}(a) as functions of the collision energy 
for several different $\tau{_{\rm F}}$ values. 
For our method (solid), the triangular time profile (dot-dashed), or
the Bjorken formula (dotted), the three curves from top to bottom
represent the results for $\tau{_{\rm F}}=$ 0.1, 0.3 and 0.9 fm/$c$, respectively. 
At high energies and a finite $\tau{_{\rm F}}$ where $\tau{_{\rm F}} \gg d_t$,
one finds that both our solution and the triangular solution
reduce to the Bjorken formula, which can be seen in
Fig.~\ref{fig:emax-snn-tauf}(a). 
Numerically we observe that the Bjorken $\epsilon^{\rm max}$ value
starts to be significantly different (by 20\% or more) from our
finite-thickness result when $\tau{_{\rm F}}/d_t \lesssim 1$ (as we
naively expect), which may be considered as the condition when the
Bjorken energy density formula breaks down.

At low energies, our $\epsilon^{\rm max}$ value is 
much smaller than that from the Bjorken formula and its dependence on
$\tau{_{\rm F}}$ is also much weaker. These qualitative features are the same
as those found in the earlier study \cite{Lin:2017lcj}. 
Furthermore, we find that numerically $\tau{_{\rm F}}/d_t \lesssim 0.2$ when our
$\epsilon^{\rm max}$ value is
significantly different (by 20\% or more) from the previous
triangular solution, and usually our $\epsilon^{\rm max}$ is smaller
than the triangular solution at very low energies but is bigger at
intermediate energies. Note that for the triangular time profile of
the initial energy production the maximum energy density is given by
\cite{Lin:2017lcj} 
\begin{eqnarray}
\epsilon^{\rm max}_{\rm tri} &&=\frac{2}{A_{\rm T} t_{21}}
\frac{dm_{\rm T}}{dy}(0) \left [ \frac{}{}
-1-\frac{\tau{_{\rm F}}}{t_{21}}+\sqrt {\frac{\tau{_{\rm F}}}{t_{21}}} \sqrt
{2 +\frac{\tau{_{\rm F}}}{t_{21}}} \right . \nonumber \\  
&& \left . +2 \ln \lefttwo \left ( \frac {1+\sqrt
{1+2\;t_{21}/\tau{_{\rm F}}}}{2} \right ) \right ]. 
\label{emaxtri}
\end{eqnarray}

We know that the energy density from the Bjorken formula
diverges as $1/\tau{_{\rm F}}$, while the triangular solution 
diverges as $1/\tau{_{\rm F}}$ at high energies but as 
$\ln{(1/\tau{_{\rm F}})}$ at low energies \cite{Lin:2017lcj}. 
Figure~\ref{fig:emax-snn-tauf}(b) shows how the maximum energy
density depends on the formation time $\tau{_{\rm F}}$, where  
solid curves show our results for central Au+Au collisions 
at $\snn=$ 3, 10 and 200 GeV as functions of $\tau{_{\rm F}}$.  
We see a flattening of $\epsilon^{\rm max}$  as $\tau{_{\rm F}}$ decreases towards
zero, which is more obvious at lower energies. 
Also, our results are close to results from the previous triangular
solution  (dot-dashed) when the formation time is not too small.  
On the other hand, energy densities from the Bjorken  
formula (dotted lines) go as $1/\tau{_{\rm F}}$ 
and are much higher than our results at low energies 
and/or small $\tau{_{\rm F}}$ values.

\section{Finiteness of $\epsilon^{\rm max}$ at $\tau{_{\rm F}}=0$}

We further find that the maximum initial energy density at $\tau{_{\rm F}}=0$, 
$\epsilon^{\rm max}(\tau{_{\rm F}}\!=\!0)$, is finite, and the
values are shown as circles in Fig.~\ref{fig:emax-snn-tauf}(b) for those
three energies.  
Note that $\epsilon^{\rm max}(\tau{_{\rm F}}\!=\!0)$ is finite at any energy,
and its energy dependence  is shown in Fig.~\ref{fig:emax-snn-tauf}(a)
as the curve with circles. 
We see that the $\epsilon^{\rm max}(\tau{_{\rm F}}=0)$ value is quite
close to (within 20\% of) the $\epsilon^{\rm max}$ value at $\tau{_{\rm F}}=0.1$ fm/$c$  
for central Au+Au collisions at $\snn \lesssim 7$ GeV.

As an analytical proof of the finiteness of $\epsilon^{\rm
  max}(\tau{_{\rm F}}=0)$, next we derive its upper bound. 
Equation~\eqref{y0} allows us to write 
\begin{equation}
z_0=-r_0 \, {\rm \sinh}\,y_0, ~t-x=r_0 \, {\rm \cosh}\,y_0
\end{equation}
for partons that contribute to the energy density $\epsilon(t)$ at
$\eta_s\approx 0$. 
For brevity we write the variable $y_0$ as $y$ in the rest of this
section, we can then write Eq.\eqref{epsilon2} as
\begin{equation}
\begin{split}
&\epsilon(t)=\frac{2}{A_{\rm T}\beta t_{21}^2} 
\iint_S \frac{dm_{\rm T}}{dy} {\rm \cosh}^2y~dr_0dy \\
&=\frac{2}{A_{\rm T}\beta t_{21}^2} 
\int\frac{dm_{\rm T}}{dy} {\rm \cosh}^2y\, \Delta r_0(y)\, dy, 
\end{split}
\label{epsilon dr0 dy}
\end{equation}
where $\Delta r_0(y) \equiv r_0^{\rm max}(y)-r_0^{min}(y)$.
By analyzing the general crossing diagram (i.e., the one using $t_1$
and $t_2$) similar to Fig.~\ref{fig:crossing-diagram}, we first find 
that for $t \leq t_2$ we always have
\begin{equation}
r_0^{min}(y) = 0, ~r_0^{\rm max}(y) \leq r_1(y),
\end{equation}
for a given parton rapidity $y$ when $\tau{_{\rm F}}=0$. 
In the above, $r_1(y)$ is the $r_0$ value when a
parton passing through the observation point $(0,t)$ with rapidity $y$
intersects one of the $z=\pm \beta (x-t_1)$ lines in the general
crossing diagram:
\begin{equation}
r_1(y)=\frac{\beta(t-t_1)}{\beta {\rm \cosh}\,y
+|  {\rm \sinh}\,y | }.
\label{r0max upper bound}
\end{equation}
Thus for $t \leq t_2$ we have
\begin{equation}
\Delta r_0(y) \leq 
\frac{\beta t_{21}}{\beta {\rm \cosh}\,y+|  {\rm \sinh}\,y | }.
\label{relation1}
\end{equation}
Secondly, for $t \geq t_2$ we can obtain
\begin{equation}
\Delta r_0(y) =\frac{\beta \, {\rm \cosh} y \, t_{21} - 
2 |{\rm \sinh} y| (t- t_{\rm mid})}{\beta {\rm \cosh}^2 y
  -{{\rm \sinh}^2 y}/\beta},
\end{equation}
which also satisfies the inequality of Eq.\eqref{relation1}.
Equation~\eqref{epsilon dr0 dy} then gives 
\begin{equation}
\epsilon(t) \leq \frac{2}{A_{\rm T}t_{21}} \frac{dm_{\rm T}}{dy}(0)
\lefttwo \int \lefttwo \frac{e\!^{^{-\frac{y^2}{2\sigma^2}}}\;{\rm
    \cosh}^2y~dy}{\beta {\rm \cosh}\,y \!+\!|  {\rm \sinh}\,y | }
\equiv \epsilon_{\rm bound}. 
\label{upperBound}
\end{equation}

\begin{figure}
\centering
\includegraphics[width=\linewidth]{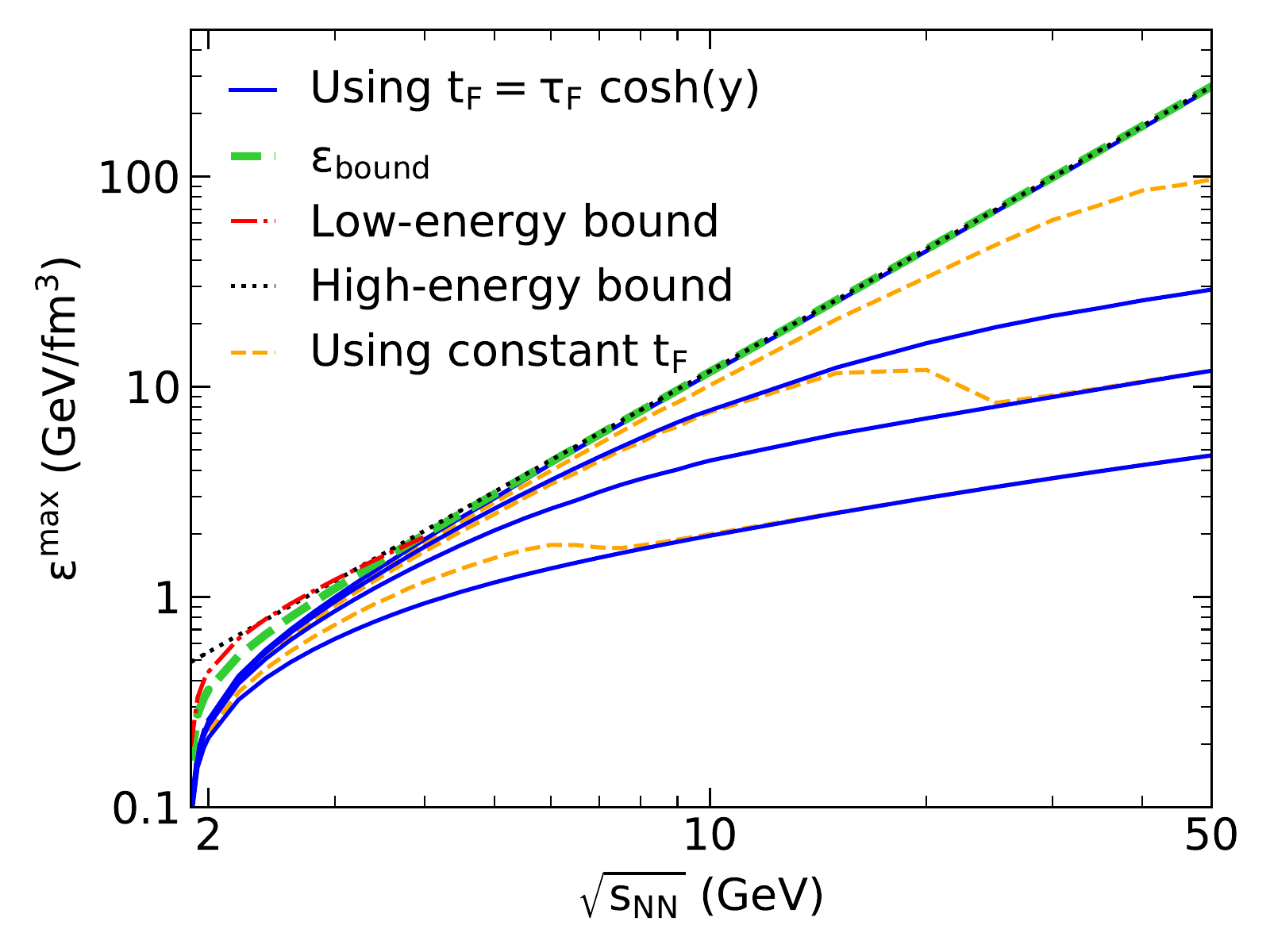}
\caption{Maximum energy density for central Au+Au collisions 
as a function of collision energy at $\tau{_{\rm F}}=0, 0.1, 0.3$ and 0.9
fm/$c$ in comparison with the upper bound of $\epsilon^{\rm max}$ of
Eq.\eqref{upperBound}, where the analytical low- and high-energy bounds are
also shown. Dashed curves represent the $\epsilon^{\rm max}$ results
when using a constant formation time $t_{\rm F}=0.1, 0.3$, and 0.9 fm$/c$.}
\label{fig:emax-bound}
\end{figure}

This upper bound of the energy density is shown (thick dashed curve)
for central Au+Au collisions as a function of the collision energy 
in Fig.~\ref{fig:emax-bound}.  
We observe that it approaches the $\epsilon^{\rm max}(\tau{_{\rm F}}=0)$ value
(the top solid curve) at high energies. Note that as $\beta \to 1$ the
light cone boundaries overlap with the upper boundaries of the rhombus 
production area, thus the inequality of Eq.\eqref{relation1} becomes
an equality for $t \leq t_2$ but not for $t>t_2$. Therefore the
observation $\epsilon_{\rm bound} \rightarrow \epsilon^{\rm
  max}(\tau{_{\rm F}}=0)$ at high energies suggests that the maximum energy
density for $\tau{_{\rm F}}=0$ is reached at $t \leq t_2$. This is the
case for the triangular time profile \cite{Lin:2017lcj}, where 
$\epsilon^{\rm max}_{\rm tri} = \epsilon_{\rm tri} (t=t_{\rm
  mid}+\tau{_{\rm F}}/2+\sqrt {\tau{_{\rm F}}} \sqrt {2\;t_{21}+\tau{_{\rm F}}}/2 )$
occurs at a time within $[t_{\rm mid}+\tau{_{\rm F}},t_2+\tau{_{\rm F}})$.

For an explicit analytical expression of the upper bound, we take
advantage of 
\begin{equation}
\frac{1}{\beta {\rm \cosh}\,y+|{\rm \sinh}\,y|} \leq \frac
{e^{-|y|}}{\beta}.
\label{relax}
\end{equation}
Using Eq.\eqref{sigma}, we then reduce Eq.\eqref{upperBound} to 
\begin{equation}
\epsilon(t) \leq\frac{A\snn}{2A_{\rm T}\beta t_{21}} 
\left [2 \! +\! \text{erfc}\! \left(\frac{\sigma}{\sqrt{2}} \right) 
\! + \! e^{4\sigma^2 }\! \text{erfc}\left( \frac{3\sigma}{\sqrt{2}}
  \right) \right ],
\label{upperBoundHigh}
\end{equation}
where $\text{erfc}(x)$ is the complementary error function. 
The right hand side of Eq.\eqref{upperBoundHigh} can be
considered as the high energy expression of the upper bound, and 
as shown in Fig.~\ref{fig:emax-bound} (dotted curve) it agrees
  well with $\epsilon_{\rm bound}$ of Eq.\eqref{upperBound} 
for $\snn > 4$ GeV.
For very low energies, however, the relaxation of 
Eq.\eqref{relax} is too loose and thus the high energy bound 
of Eq.\eqref{upperBoundHigh} fails to approach zero at
the threshold energy.

At very low energies where $\beta \ll 1$, 
we find from Eq.\eqref{sigma} that $\sigma < 0.707$ for $\snn<1.96$
GeV. Using the fact $\exp(-y^2/2/\sigma^2)\,{\rm \cosh}^2y \leq 1$ for
$\sigma < 1/\sqrt{2}$, Eq.\eqref{upperBound} gives  
\begin{equation}
\begin{split}
&\epsilon(t) \leq
\frac{2}{A_{\rm T}t_{21}} \frac{dm_{\rm T}}{dy}(0)
\int \frac{dy}{\beta {\rm \cosh}\,y+|  {\rm \sinh}\,y | }  \\
& = \frac{8}{A_{\rm T}t_{21}\sqrt{1-\beta^2}} \frac{dm_{\rm T}}{dy}(0)
\tanh^{-1} \!\! \left ( \sqrt {\frac{1-\beta}{1+\beta}} \right ).  
\end{split}
\label{upperBoundLow}
\end{equation}
This low energy expression of the upper bound is shown in 
Fig.~\ref{fig:emax-bound} 
(dot-dashed curve), where we see that it captures the
decrease of the energy density $\epsilon^{\rm max}(\tau{_{\rm F}}=0)$ towards
the threshold energy. 
Note that $\epsilon_{\rm bound} \propto \beta \ln(2/\beta)$ at very low
energies according to Eq.\eqref{upperBoundLow}, therefore the peak
energy density goes towards zero as the collision energy approaches
the threshold although the initial transverse mass rapidity
density of Eq.\eqref{dmtdy0} is always finite.

\section{Scaling and $A$-dependence of $\epsilon(t)$ }

Our solution of Eq.\eqref{epsilon2} has an approximate scaling
property. We first note that, in the hard sphere model of the
nucleus, both the time duration $d_t$ and the $z$-width of the
production area are proportional to $A^{1/3}$. 
Secondly, we can expect $dm_{\rm T}/dy(0)$ to be
approximately proportional to the number of participant nucleons 
and thus proportional to $A$ for central collisions; this is the case
for the parametrization of the final hadron $dE_{\rm T}/dy$ by the
PHENIX Collaboration \cite{Adler:2004zn}. 
If $dm_{\rm T}/dy(0) \propto A$ for central collisions, 
Eq.\eqref{sigma} means that the Gaussian width $\sigma$ of the
$dm_{\rm T}/dy$ distribution is independent of $A$. 

Next we define the scaled time and scaled proper formation time
respectively as 
\begin{equation}
t^s=\frac{t}{A^{1/3}},~\tau{_{\rm F}^s}=\frac{\tau{_{\rm F}}}{A^{1/3}}.
\end{equation}
Under these approximations (i.e., $d_t$ and the $z$-width of the
production area are proportional to $A^{1/3}$ and $dm_{\rm T}/dy(0)
\propto A$), we see from Eq.\eqref{epsilon2} that at a given collision
energy $\epsilon(t)$ is only a function of $t^s$ and $\tau{_{\rm F}^s}$,  
while $\epsilon^{\rm max}$ is only a function of $\tau{_{\rm F}^s}$. 
This also gives the following scaling relation:
\begin{equation}
\epsilon^{\rm max}_{\rm AA}({\rm for~} \tau{_{\rm F}})=
\epsilon^{\rm max}_{\rm AuAu} \left ( {\rm for~} \tau{_{\rm F}}^{\rm Au} \!=\!
(197/A)^{1/3} \tau{_{\rm F}} \right)
\label{scaling}
\end{equation}
at the same energy ($\snn$). For example, it means 
$\epsilon^{\rm max}_{\rm OO}({\rm for~} \tau{_{\rm F}}=0.30{\rm~fm}/c)=
\epsilon^{\rm max}_{\rm AuAu}({\rm for~} \tau{_{\rm F}}=0.69{\rm~fm}/c)$
for central collisions at the same energy. In addition, it means that
$\epsilon^{\rm   max}_{\rm AA}(\tau{_{\rm F}}=0)$ only depends on $\snn$ but
not on $A$.  If one were willing to apply these approximations down to
$A=1$ (for the proton), Eq.\eqref{scaling} would give $\epsilon^{\rm
  max}_{\rm   AA}(\tau{_{\rm F}}=0)=\epsilon^{\rm max}_{\rm pp}(\tau{_{\rm F}}=0)$  
for central $AA$ collisions at the same energy.

Furthermore, the scaling means that the $\tau{_{\rm F}}$-dependence
of $\epsilon^{\rm max}$ at a given energy, such as the curves shown in
Fig.~\ref{fig:emax-snn-tauf}(b), also gives the $A$-dependence of
$\epsilon^{\rm max}$ for central collisions. 
We see that the Bjorken formula Eq.\eqref{bjorken}
and the triangular solution Eq.\eqref{ettri} 
also satisfy the scaling relation. 
However, different $\tau{_{\rm F}}$-dependences correspond to  
different $A$-dependences of the maximum energy density. 
For example, at low energies our result has a 
very flat $\tau{_{\rm F}}$-dependence as shown in 
Fig.~\ref{fig:emax-snn-tauf}(b), which translates to a 
very slow increase of $\epsilon^{\rm max}$ with $A$. 
At finite $\tau{_{\rm F}}$ and high-enough energies, however, 
our result reduces to the Bjorken energy density formula, 
where $\epsilon^{\rm max} \propto A^{1/3}$ at fixed $\tau{_{\rm F}}$. 
Also note that under the same approximations the upper bound of the
energy density  $\epsilon_{\rm bound}$ in Eq.\eqref{upperBound} is 
independent of $A$, just like $\epsilon^{\rm max}_{\rm AA}(\tau{_{\rm F}}=0)$. 

\section{Discussions}

In the calculations of energy density with Eq.\eqref{epsilon2} so far, 
we have taken $dm_{\rm T}/dy$ as the transverse mass rapidity
density of initial partons, which peak value as a function of energy is
parametrized according to results from the AMPT model
\cite{Lin:2004en,Lin:2017lcj}.  
To investigate the uncertainty of the energy density due to $dm_{\rm
  T}/dy$,  we could also take $dm_{\rm T}/dy$ as the transverse mass 
rapidity density of final hadrons. 
The hadron $dm_{\rm T}/dy$ is derived in the Appendix and shown in
Fig.~\ref{fig:dmtdy} (dashed curves) for central Au+Au collisions at 
several energies.  
We see that the hadron $dm_{\rm T}/dy$ and parton $dm_{\rm T}/dy$ are
similar at energies between $\approx 3$ and 50 GeV.
At 2 GeV near the threshold energy, however, the hadron $dm_{\rm
  T}/dy$ has a higher peak but is narrower than the parton $dm_{\rm 
  T}/dy$ because of the slow baryons, while the hadron
$dm_{\rm T}/dy$ has a lower peak at the top RHIC energy consistent
with the effect of strong secondary interactions. 
Note that both the hadron and parton $dm_{\rm T}/dy$ satisfy the
energy conservation of Eq.\eqref{energy-conservation}.

\begin{figure}
\centering
\includegraphics[width=\linewidth]{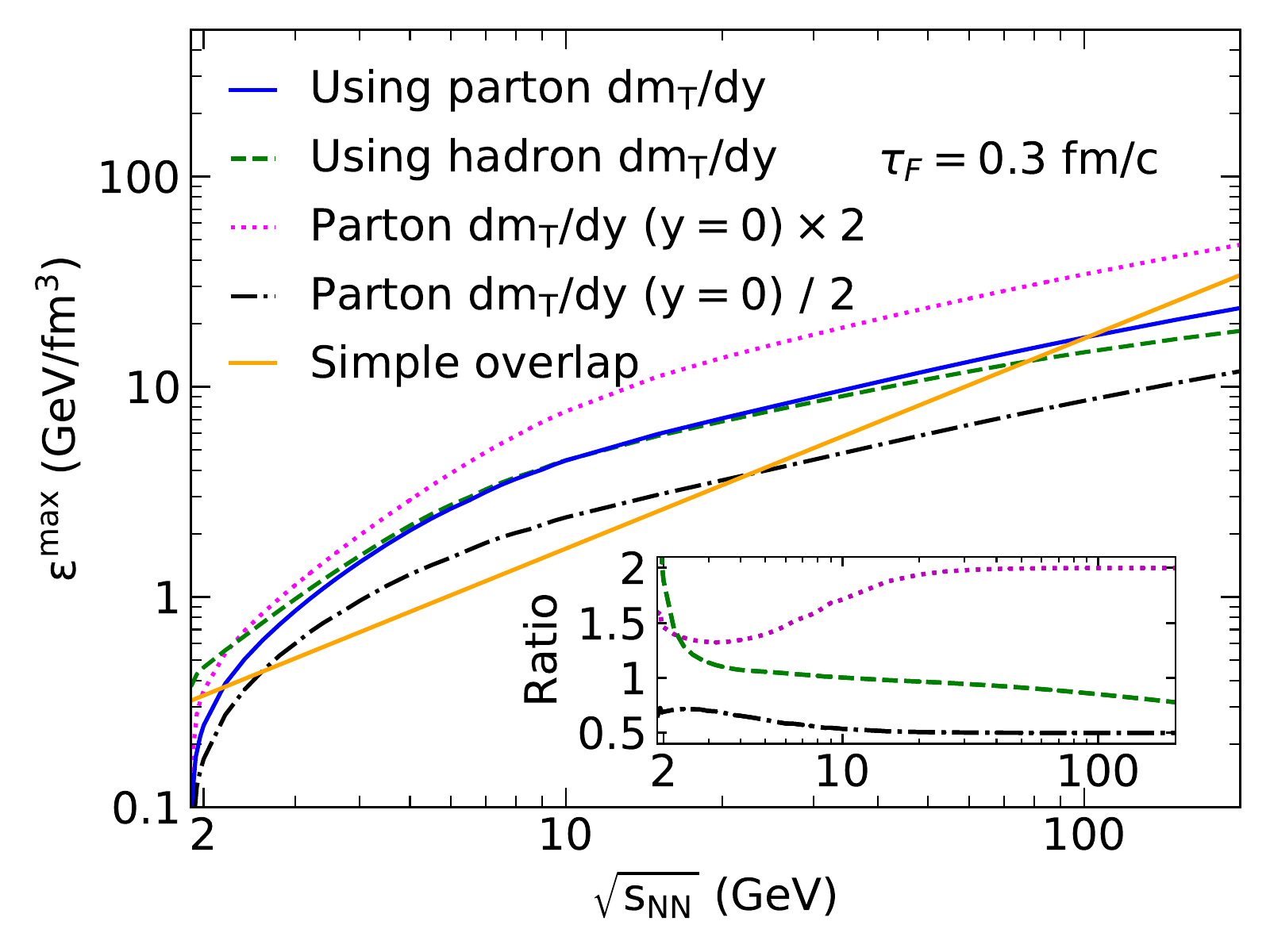}
\caption{Maximum energy density using the parton
$dm_{\rm T}/dy$, the hadron $dm_{\rm T}/dy$, 
or modified parton $dm_{\rm T}/dy$ (see text for details)
for $\tau{_{\rm F}}=$ 0.3 fm/$c$ as functions of energy. 
The inset shows ratios of $\epsilon^{\rm max}$ from various
$dm_{\rm T}/dy$ profiles to $\epsilon^{\rm max}$ from the default
parton $dm_{\rm T}/dy$, while the solid straight line shows the energy
density if two boosted nuclei simply overlap.}
\label{fig:emax-parton-hadron}
\end{figure}

Figure~\ref{fig:emax-parton-hadron}
shows the $\epsilon^{\rm max}$ values in central Au+Au collisions 
as functions of energy when the hadron $dm_{\rm T}/dy$ is used (dashed
curve) for $\tau{_{\rm F}}=0.3$ fm/$c$.  
We see that it is rather close to our result for the parton $dm_{\rm
  T}/dy$ (solid curve) within $3 < \snn < 100$ GeV. 
At high energies the production area is relatively small 
compared to the finite $\tau{_{\rm F}}$, so partons with rapidities near zero
dominate the energy at $\eta_s\approx 0$. 
Therefore $\epsilon^{\rm max}$ at high collision energies is expected
to depend mostly on $dm_{\rm T}/dy(0)$; as a result, 
the $\epsilon^{\rm max}$ value using the hadron $dm_{\rm T}/dy$ is 
lower than that using the parton $dm_{\rm T}/dy$ (see
Fig.~\ref{fig:dmtdy}). At low energies particles at finite rapidities
can also contribute significantly to the energy  at $\eta_s\approx 0$,
thus $\epsilon^{\rm max}$ depends on not only $dm_{\rm T}/dy(0)$ but
also the Gaussian width $\sigma$. 
To further demonstrate this, we have changed the parton $dm_{\rm
  T}/dy(0)$ value by a factor of 2 and then determined the Gaussian
width with the energy conservation of Eq.\eqref{energy-conservation};
the corresponding $\epsilon^{\rm max}$ values are shown in 
Fig.~\ref{fig:emax-parton-hadron} with the ratio over our default
result (solid curve) shown in the inset. 
We see that the change of $\epsilon^{\rm max}$ is the same factor
of 2 at high energies but is smaller than two at low energies.

We also show in Fig.~\ref{fig:emax-parton-hadron} the simplest estimate
for the energy density (straight line), where one imagines the two
boosted nuclei to simply overlap in volume with all interactions
neglected. In the hard sphere model of the nucleus, this energy
density would be 
\begin{equation}
\epsilon^{\rm overlap}=\frac{3\snn}{4\pi R_1^3},
\end{equation}
which grows linearly with $\snn$ but is independent of $A$. 
Naively we expect the actual maximum energy density in the central
spacetime-rapidity region to be higher than $\epsilon^{\rm
  overlap}$ due to the compression from the primary nucleus-nucleus
collision. This is indeed the case in Fig.~\ref{fig:emax-parton-hadron}
except for very low or very high energies.  
Near the threshold energy the energy density using the hadron $dm_{\rm
  T}/dy$ is higher than $\epsilon^{\rm   overlap}$, but the energy
density using the parton $dm_{\rm T}/dy$ is lower. 
However, a parton matter is unlikely to be formed near the threshold
energy due to the low estimated energy density, therefore the hadron
$dm_{\rm T}/dy$ should be more applicable there. 
At very high energies, we expect the parton $dm_{\rm T}/dy$ to be
applicable but the maximum energy density is lower than 
$\epsilon^{\rm overlap}$. This is because of the finite formation time
$\tau{_{\rm F}}$; for example we see from Fig.~\ref{fig:emax-snn-tauf}(a) that
the peak energy density at $\tau{_{\rm F}}=0$ at high energies is always bigger
than $\epsilon^{\rm overlap}$.

We have also considered a scenario where all initial partons 
have the same formation time $t_{\rm  F}$ instead of 
the same proper formation time $\tau{_{\rm F}}$.
The energy density is still given by Eqs.\eqref{epsilon} and
\eqref{epsilon2}, but the formation time requirement restricts the
integration area $S$ to $x \leq t-t_{\rm  F}$ instead of restricting $S$
below the proper time hyperbola of
Eq.\eqref{tauFcurve}. Figure~\ref{fig:emax-bound} shows the
$\epsilon^{\rm max}$ results (thin dashed curves) for $t_{\rm F}=0.1,
0.3$ and 0.9 fm$/c$ as functions of energy, where the result above a
certain energy (which corresponds to $\beta t_{21}/2 \approx t_{\rm  F}$)
is the same as our standard result that takes the same value for
$\tau{_{\rm F}}$. However, just below this energy scale we see a strange
decrease of $\epsilon^{\rm max}$ with $\snn$. 
We find that this is a consequence of a double-peak structure of 
$\epsilon(t)$ below this energy scale in the constant-$t_{\rm F}$
case, where partons at very large rapidities could also contribute to
the energy density. 

\section{Conclusion}

We present a method to calculate the initial energy density
produced in heavy ion collisions that takes into account 
the finite nuclear thickness. 
Our method includes both the finite longitudinal ($z$-) width and the
finite time duration $d_t$ of the initial energy production. 
This is a continuation of a previous study that
considers the finite duration time (but 
not the finite $z$-width) in an extension of the Bjorken energy
density formula. 
We find the same qualitative conclusions: 
the initial energy density after considering the finite nuclear
thickness approaches the Bjorken formula at large formation time
$\tau{_{\rm F}}$ and/or high energies; at low energies, however, the initial 
energy density has a much lower maximum, evolves much longer, and is
much less sensitive to $\tau{_{\rm F}}$ than the Bjorken formula. 
Numerically we find that the Bjorken energy density formula breaks
down (i.e., is different by 20\% or more from our results that include
the finite nuclear thickness) when $\tau{_{\rm F}}/d_t \lesssim 1$, as one
may expect. When the proper formation time $\tau{_{\rm F}}$ is
not too much smaller than the crossing time of the two nuclei, 
our results are similar to the previous extension
results that only include the finite time duration. Numerically we
find $\tau{_{\rm F}}/d_t \lesssim 0.2$ when our result is
significantly different (by 20\% or more) from the  previous result.  

A qualitative difference from previous studies is that we find the
energy density $\epsilon(t)$ including its maximum $\epsilon^{\rm
  max}$ to be finite at $\tau{_{\rm F}}=0$ at any energy.   
In contrast, the Bjorken energy density formula is divergent where
$\epsilon^{\rm max} \propto 1/\tau{_{\rm F}}$ as $\tau{_{\rm F}} \to 0$, 
while the previous study that neglects the finite $z$-width
gives a $\ln (1/\tau{_{\rm F}})$ divergence at low energies but  
the same $1/\tau{_{\rm F}}$ divergence at high energies. 

In addition, we find that our $\epsilon(t)$ results (as well as the
Bjorken energy density formula and the previous extension results)  
for central heavy ion collisions satisfy a scaling relation under 
two reasonable assumptions. They include the assumption 
that the initial rapidity density of the transverse energy
is proportional to the number of participant nucleons and that the
$z-$width and time duration $d_t$ are both proportional to  
$A^{1/3}$. As a result of the scaling, the $\tau{_{\rm F}}$-dependence of
$\epsilon^{\rm max}$ for a given $A$ also determines the
$A$-dependence of $\epsilon^{\rm max}$ (at the same collision energy),
therefore the weaker $\tau{_{\rm F}}$-dependence of our results at low 
energies means a slower increase of the energy density with the mass
number $A$. In particular, the scaling means that the $\epsilon^{\rm
  max}$ value at $\tau{_{\rm F}}=0$ is independent of $A$ and only 
depends on the collision energy.

\appendix

\section{$dm_{\rm T}/dy$ of final state hadrons}
\label{hadron-dmtdy}

In the PHENIX Collaboration's data-based parametrization
\cite{Adler:2004zn} of the transverse energy pseudo-rapidity density
around $\eta=0$, the ``transverse energy'' $E_{\rm T}$ is defined as 
$E_{\rm T}=\sum_iE_i\sin{\theta_i}$, 
where $\theta_i$ is the polar angle of particle $i$. 
$E_i$ is defined as $E_i^{tot}-m_{\rm N}$ for baryons,
$E_i^{tot}+m_{\rm N}$ for antibaryons, and $E_i^{tot}$ for all other
particles, where  $E_i^{tot}$ is the total energy of the particle and
$m_{\rm N}$ is the nucleon mass. As a result of the $E_{\rm T}$
definition, the total transverse energy of hadrons at $y=0$ is given
by
\begin{equation}
\frac{dm_{\rm T}}{dy}=\frac{dE_{\rm T}}{dy}+m_{\rm N}\frac{dN_{\rm netB}}{dy},
\label{hadron-dmtdy}
\end{equation}
where $N_{\rm netB}$ represents the net-baryon number. 

To determine the hadron $dm_{\rm T}/dy$ function for 
calculating the energy density via Eq.\eqref{epsilon2}, 
we assume that $dE_{\rm T}/dy$ is a single Gaussian 
while $dN_{\rm netB}/dy$ can be described with a double-Gaussian
\cite{Anticic:2003ux,MehtarTani:2008qg}:
\begin{equation}
\begin{split}
&\frac{dE_{\rm T}}{dy}=\frac{dE_{\rm T}}{dy}(0)~{\mathlarger
  e}^{-\frac{y^2}{2\sigma_1^2}},\\ 
&\frac{dN_{\rm netB}}{dy}=C \left ( {\mathlarger e}^{-\frac{(y+{y_B})^2}{2\sigma_2^2}}
+ {\mathlarger e}^{-\frac{(y-{y_B})^2}{2\sigma_2^2}} \right ).
\end{split}
\label{gaussian}
\end{equation}

\begin{figure}
\begin{minipage}[h]{\linewidth}
\includegraphics[width=1.\textwidth]{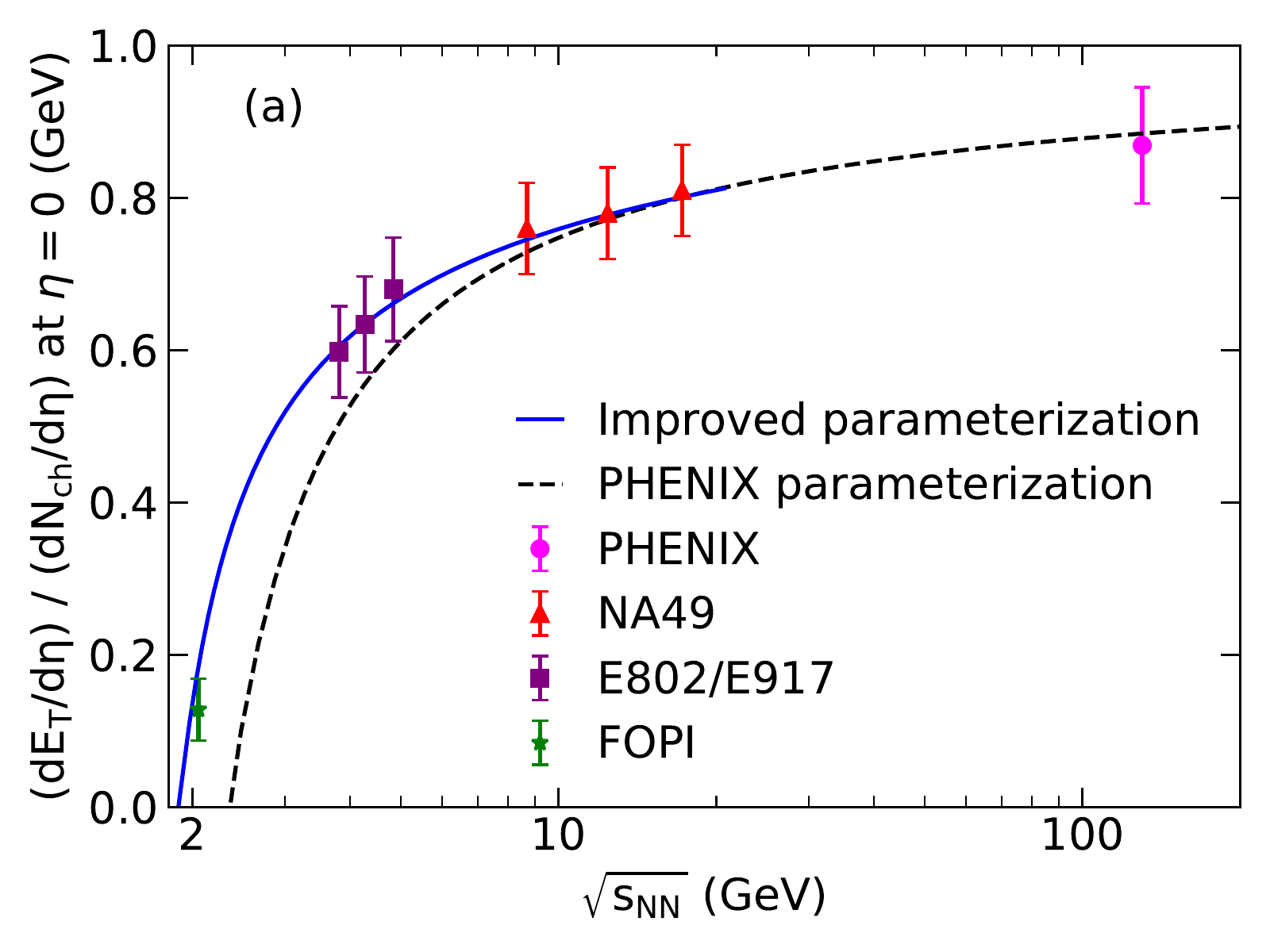}
\end{minipage}
\hfill
\begin{minipage}[h]{\linewidth}
\includegraphics[width=1.\textwidth]{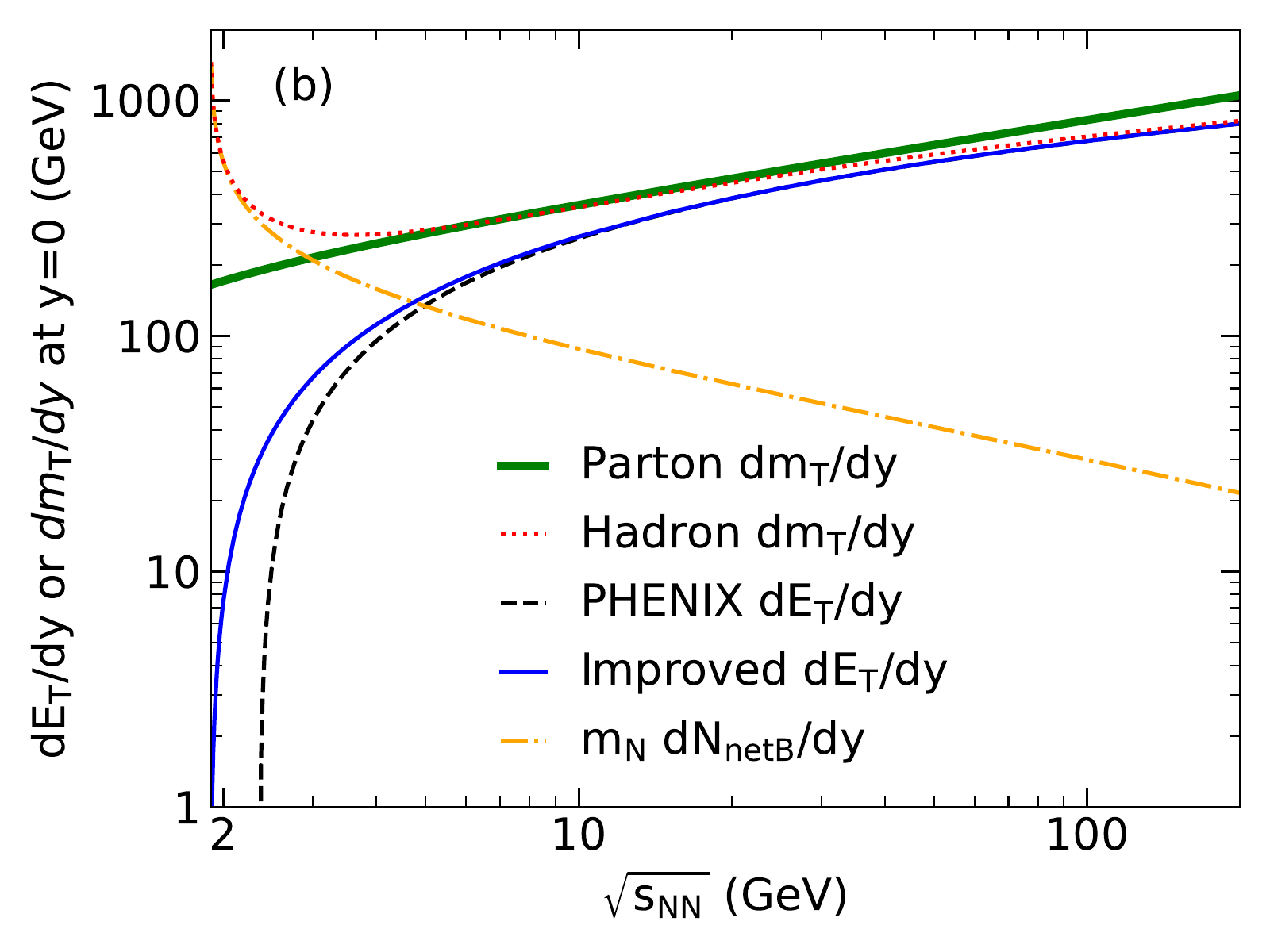}
\end{minipage}
\caption{(a) $(dE_{\rm T}/d\eta)/(dN_{\rm ch}/d\eta)$ data 
at $\eta \approx 0$ compared with our improved parametrization 
below 20.7 GeV and the PHENIX parametrization.
(b) Mid-rapidity $dm_{\rm T}/dy$ of initial partons 
and final hadrons for central Au+Au
collisions as functions of energy; the net-baryon contribution 
to the hadron $dm_{\rm T}/dy\,(0)$ as well as the PHENIX
parametrization and our improved parametrization of
$dE_{\rm T}/dy\,(0)$ are also shown.}
\label{fig:detdy}
\end{figure}

First, regarding $dE_{\rm T}/dy(0)$ the PHENIX Collaboration has
parametrized the mid-pseudorapidity data as \cite{Adler:2004zn}
\begin{equation}
\begin{split}
&\frac {dN_{\rm ch}}{d\eta}(0)=0.37N_p \ln \left
  (\frac{\snn}{1.48{\rm GeV}} \right ),\\
&\frac {dE_{\rm T}}{d\eta}(0)=0.365N_p\ln \left
  (\frac{\snn}{2.35{\rm GeV}} \right ) {\rm ~GeV},\\
\end{split}
\label{phenix}
\end{equation}
where $N_p$ is the number of participants 
(taken as $2A$ for central collisions in this study). 
However, the $dE_{\rm T}/d\eta(0)$ parametrization underestimates the 
$dE_{\rm T}/d\eta/(dN_{\rm ch}/d\eta)$ ratio at energies below $\snn
\approx 10$ GeV \cite{Adler:2004zn}, as shown in Fig.~\ref{fig:detdy}(a).
Since the effect of finite
nuclear thickness is more important at lower energies and the 
PHENIX parametrization of $dN_{\rm ch}/d\eta(0)$ is accurate 
down to lower energies than that of $dE_{\rm T}/d\eta(0)$, 
we improve the $dE_{\rm T}/d\eta(0)$ parametrization. 
Specifically, we take the same $dN_{\rm ch}/d\eta(0)$ parametrization
\cite{Adler:2004zn} but refit the $dE_{\rm T}/d\eta/(dN_{\rm
  ch}/d\eta)$ data at $\snn < 20$ GeV to obtain
\begin{equation}
\frac{dE_{\rm T}}{d\eta}(0) =
0.308 N_p \ln^{1.08} \left (\frac{\snn}{E_0}\right) {\rm ~GeV},
\end{equation}
for $\snn \leq 20.7$ GeV, where $E_0=2m_{\rm N}$ is the threshold energy.
As shown in Fig.~\ref{fig:detdy}(a), our improved low energy parametrization
intersects the PHENIX parametrization at $\snn \approx 20.7$ GeV, above
which we use the PHENIX $dE_{\rm T}/d\eta(0)$ parametrization. We
then take $dE_{\rm T}/dy(0)=1.25\ dE_{\rm T}/d\eta(0)$
\cite{Adler:2004zn}, which are shown in Fig.~\ref{fig:detdy}(b) 
for our improved parametrization (thin solid curve) 
and the PHENIX parametrization (dashed curve).

\begin{figure}
\includegraphics[width=\linewidth]{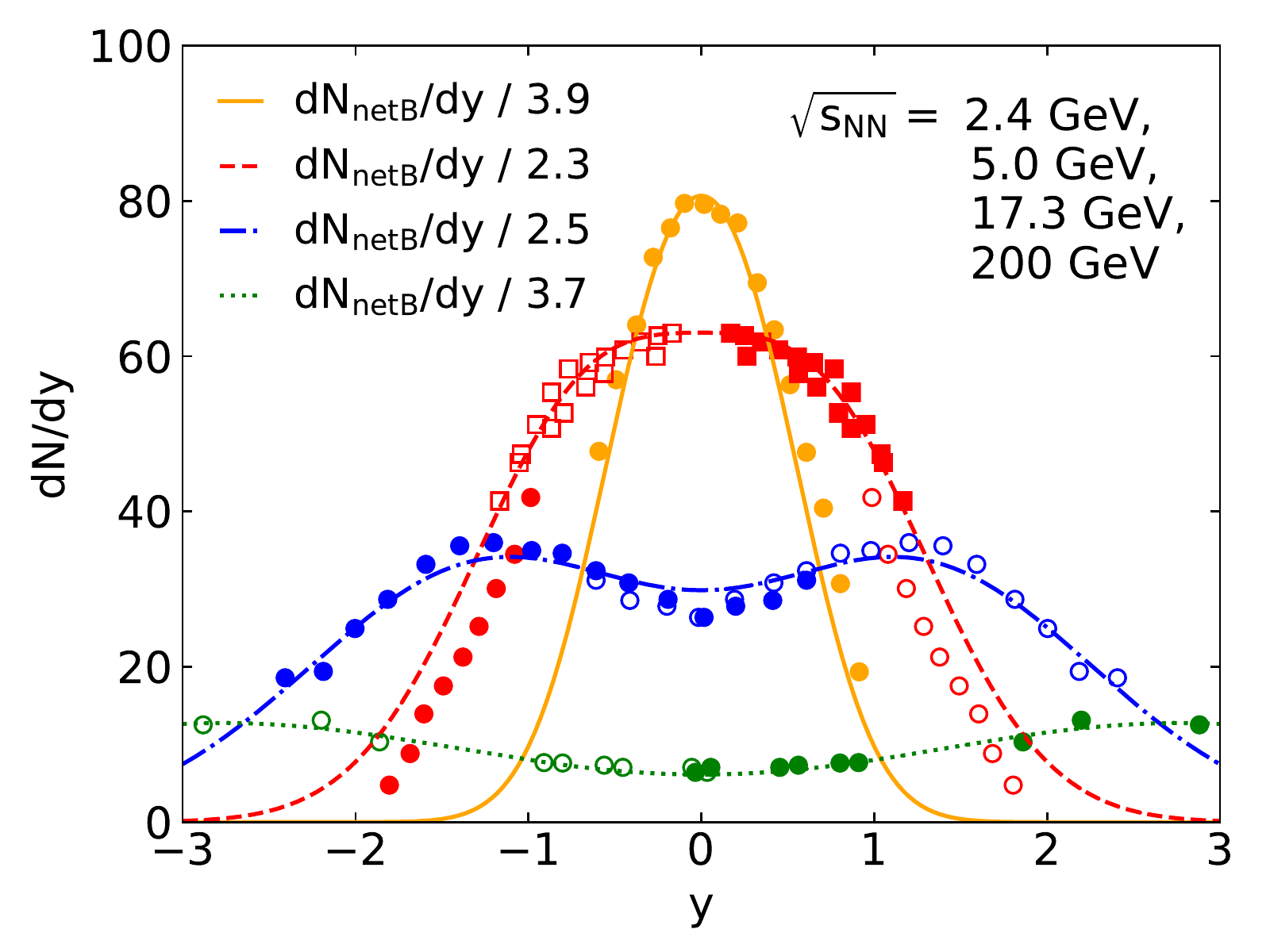} 
\caption{Net-proton $dN/dy$ data (circles) for central Au+Au
  (Pb+Pb) at $\snn=$ 2.4, 5, (17.3), and 200 GeV in comparison with 
the scaled net-baryon parametrization (curves). Filled circles
represent actual data and open circles are reflected data across $y=0$.
}
\label{fig:netp}
\end{figure}

Next, to specify $dN_{\rm netB}/dy$ in Eq.\eqref{gaussian} we first
parametrize $y_B$ and $\sigma_2$ using the net-proton rapidity
density data in central Au+Au collisions (with the exception that 
central Pb+Pb data are used at 17.3 GeV). 
For collision energies below 5 GeV, there is little anti-baryon
production and thus we use the proton $dN/dy$ for net-protons at
$\snn$ = 2.4, 3.1, 3.6, and 4.1 GeV \cite{Klay:2001tf}.  
We also use the net-proton $dN/dy$ data at $\snn=5$ GeV
\cite{Ahle:1999in,Barrette:1999ry}, 17.3 GeV
\cite{Appelshauser:1998yb} and 200 GeV \cite{Bearden:2003hx}.  
From these data we obtain the following parametrization:
\begin{equation}
\begin{split}
&y_B=0.541 \left( \frac {\snn-E_0}{\rm GeV} \right)^{0.196} \ln^{0.392} \left (\frac{\snn}{E_0}
\right),\\
&\sigma_2=0.601 \left( \frac {\snn-E_0}{\rm GeV} \right)^{0.121} \ln^{0.241} \left
  (\frac{\snn}{E_0} \right).
\end{split}
\end{equation}
We further assume that the net-baryon and net-proton
$dN/dy$ distributions have the same shape. 
We then impose the conservation of the net-baryon number, 
$\int (dN_{\rm netB}/dy) dy=2A$ to determine the parameter $C$ in
Eq.\eqref{gaussian} at each collision energy. 
Figure~\ref{fig:netp} shows the net-proton data at several energies  
in comparison with our $dN_{\rm netB}/dy$ parametrization (scaled down by
various factors for better comparison of the shapes).  
Note that the 5 GeV data shown in Fig.~\ref{fig:netp} include those from
the E802 Collaboration (squares) \cite{Ahle:1999in} and 
the E877 Collaboration (circles) \cite{Barrette:1999ry}. 
Lastly, we calculate the last parameter $\sigma_1$ in Eq.\eqref{gaussian}
by using the conservation of total energy of
Eq.\eqref{energy-conservation}.

Figure~\ref{fig:detdy}(b) shows the energy dependence of our hadron
$dm_{\rm T}/dy(0)$ parametrization (dotted curve) in comparison with
that of the $dm_{\rm T}/dy(0)$ for initial partons (thick solid
curve). We see that they are rather close within $3 < \snn < 100$ GeV,
which includes the energy range of the Beam Energy Scan program at
RHIC
\cite{Mohanty:2011nm,Luo:2017faz,Adamczyk:2017iwn,Keane:2017kdq}. 
Note the fast increase of hadron $dm_{\rm T}/dy(0)$ when
$\snn$ decreases towards the threshold energy; this is a combined
effect of the vanishing beam rapidity near the threshold energy and
the finite conserved net-baryon number. 
It is also clear that at very low energies the net-baryon
contribution (dot-dashed curve), coming mostly from the incoming
nucleons, dominates the total transverse energy of final hadrons. 

\begin{acknowledgments}
This work has been partially supported by the National Science
Foundation under Grant No. 2012947.
\end{acknowledgments}


\begin{thebibliography}{99}

\bibitem{Gyulassy:2004zy}
M.~Gyulassy and L.~McLerran,
Nucl. Phys. A \textbf{750}, 30-63 (2005).

\bibitem{Arsene:2004fa}
I.~Arsene \textit{et al.} [BRAHMS Collaboration],
Nucl. Phys. A \textbf{757}, 1-27 (2005).

\bibitem{Back:2004je}
B.~B.~Back, \textit{et al.} [PHOBOS Collaboration],
Nucl. Phys. A \textbf{757}, 28-101 (2005).

\bibitem{Adams:2005dq}
J.~Adams \textit{et al.} [STAR Collaboration],
Nucl. Phys. A \textbf{757}, 102-183 (2005).

\bibitem{Adcox:2004mh}
K.~Adcox \textit{et al.} [PHENIX Collaboration],
Nucl. Phys. A \textbf{757}, 184-283 (2005).

\bibitem{Mohanty:2011nm}
B.~Mohanty [STAR Collaboration],
J. Phys. G \textbf{38}, 124023 (2011).

\bibitem{Luo:2017faz}
X.~Luo and N.~Xu,
Nucl. Sci. Tech. \textbf{28}, 112 (2017).

\bibitem{Adamczyk:2017iwn}
L.~Adamczyk \textit{et al.} [STAR Collaboration],
Phys. Rev. C \textbf{96}, 044904 (2017).

\bibitem{Keane:2017kdq}
D.~Keane,
J. Phys. Conf. Ser. \textbf{878}, 012015 (2017).

\bibitem{Stephanov:2011pb}
M.~A.~Stephanov,
Phys. Rev. Lett. \textbf{107}, 052301 (2011).

\bibitem{Bzdak:2019pkr}
A.~Bzdak, S.~Esumi, V.~Koch, J.~Liao, M.~Stephanov and N.~Xu,
Phys. Rept. \textbf{853}, 1-87 (2020).

\bibitem{Li:2018ygx}
Z.~Li, K.~Xu, X.~Wang and M.~Huang,
Eur. Phys. J. C \textbf{79}, 245 (2019).

\bibitem{Okai:2017ofp}
M.~Okai, K.~Kawaguchi, Y.~Tachibana and T.~Hirano,
Phys. Rev. C \textbf{95}, 054914 (2017).

\bibitem{Shen:2017ruz}
C.~Shen, G.~Denicol, C.~Gale, S.~Jeon, A.~Monnai and B.~Schenke,
Nucl. Phys. A \textbf{967}, 796-799 (2017).

\bibitem{Du:2018mpf}
L.~Du, U.~Heinz and G.~Vujanovic,
Nucl. Phys. A \textbf{982}, 407-410 (2019).

\bibitem{Bjorken:1982qr}
J.~D.~Bjorken,
Phys. Rev. D \textbf{27}, 140-151 (1983).

\bibitem{Lin:2017lcj}
Z.~W.~Lin,
Phys. Rev. C \textbf{98}, 034908 (2018).

\bibitem{Lin:2014tya}
Z.~W.~Lin,
Phys. Rev. C \textbf{90}, 014904 (2014).

\bibitem{Xu:2004mz}
Z.~Xu and C.~Greiner,
Phys. Rev. C \textbf{71}, 064901 (2005).

\bibitem{Kajantie:1983ia}
K.~Kajantie, R.~Raitio and P.~V.~Ruuskanen,
Nucl. Phys. B \textbf{222}, 152-188 (1983).

\bibitem{Spieles:1999kp}
C.~Spieles, R.~Vogt, L.~Gerland, S.~A.~Bass, M.~Bleicher, H.~St\"ocker and W.~Greiner,
Phys. Rev. C \textbf{60}, 054901 (1999).

\bibitem{Adler:2004zn}
S.~Adler \textit{et al.} [PHENIX Collaboration],
Phys. Rev. C \textbf{71}, 034908 (2005).

\bibitem{Wang:1991hta}
X.~N.~Wang and M.~Gyulassy,
Phys. Rev. D \textbf{44}, 3501-3516 (1991).

\bibitem{Gyulassy:1994ew}
M.~Gyulassy and X.~N.~Wang,
Comput. Phys. Commun. \textbf{83}, 307 (1994).

\bibitem{Lin:2004en}
Z.~W.~Lin, C.~M.~Ko, B.~A.~Li, B.~Zhang and S.~Pal,
Phys. Rev. C \textbf{72}, 064901 (2005).

\bibitem{Anticic:2003ux}
T.~Anticic \textit{et al.} [NA49 Collaboration],
Phys. Rev. Lett. \textbf{93}, 022302 (2004).

\bibitem{MehtarTani:2008qg}
Y.~Mehtar-Tani and G.~Wolschin,
Phys. Rev. Lett. \textbf{102}, 182301 (2009).

\bibitem{Klay:2001tf}
J.~Klay \textit{et al.} [E895 Collaboration],
Phys. Rev. Lett. \textbf{88}, 102301 (2002).

\bibitem{Ahle:1999in}
L.~Ahle \textit{et al.} [E802 Collaboration],
Phys. Rev. C \textbf{60}, 064901 (1999).

\bibitem{Barrette:1999ry}
J.~Barrette \textit{et al.} [E877 Collaboration],
Phys. Rev. C \textbf{62}, 024901 (2000).

\bibitem{Appelshauser:1998yb}
H.~Appelshauser \textit{et al.} [NA49 Collaboration],
Phys. Rev. Lett. \textbf{82}, 2471-2475 (1999).

\bibitem{Bearden:2003hx}
I.~Bearden \textit{et al.} [BRAHMS Collaboration],
Phys. Rev. Lett. \textbf{93}, 102301 (2004).

\end{thebibliography}
\end{document}